\documentclass[aps,prl,superscriptaddress,twocolumn]{revtex4-1}
\usepackage{amsmath}
\usepackage{epsfig}
\usepackage{graphicx}
\usepackage{color}
\usepackage{multirow}
\usepackage{array}
\usepackage{tabularx}
\usepackage{inputenc}
\newcommand{\rt}{R$_2$T$_2$O$_7$}
\newcommand{\erti}{Er$_2$Ti$_2$O$_7$}
\newcommand{\ersn}{Er$_2$Sn$_2$O$_7$}
\newcommand{\ybti}{Yb$_2$Ti$_2$O$_7$}
\newcommand{\er}{Er$^{3+}$}
\newcommand{\gd}{Gd$^{3+}$}
\newcommand{\gdsn}{Gd$_2$Sn$_2$O$_7$}
\newcommand{\gdti}{Gd$_2$Ti$_2$O$_7$}
\newcommand{\ndzr}{Nd$_2$Zr$_2$O$_7$}
\newcommand{\ybsn}{Yb$_2$Sn$_2$O$_7$}

\newcolumntype{M}[1]{>{\centering\arraybackslash}m{#1}}
\newcolumntype{Y}{>{\centering\arraybackslash}X}

\begin{document}
\title{Long range order in the dipolar XY antiferromagnet \ersn}

\author{S. Petit}
\email[]{sylvain.petit@cea.fr}
\affiliation{Laboratoire L\'eon Brillouin, CEA, CNRS, Universit\'e Paris-Saclay, CE-Saclay, 91191 Gif-sur-Yvette, France}
\author{E. Lhotel}
\email[]{elsa.lhotel@neel.cnrs.fr}
\affiliation{Institut N\'eel, CNRS and Universit\'e Grenoble Alpes, 38042 Grenoble, France}
\author{F. Damay}
\affiliation{Laboratoire L\'eon Brillouin, CEA, CNRS, Universit\'e Paris-Saclay, CE-Saclay, 91191 Gif-sur-Yvette, France}
\author{P. Boutrouille}
\affiliation{Laboratoire L\'eon Brillouin, CEA, CNRS, Universit\'e Paris-Saclay, CE-Saclay, 91191 Gif-sur-Yvette, France}
\author{A. Forget}
\affiliation{Service de Physique de l'Etat condens\'e, CEA, CNRS, Universit\'e Paris-Saclay, CE-Saclay, 91191 Gif-sur-Yvette, France}
\author{D. Colson}
\affiliation{Service de Physique de l'Etat condens\'e, CEA, CNRS, Universit\'e Paris-Saclay, CE-Saclay, 91191 Gif-sur-Yvette, France}

\begin{abstract}
\ersn\ remains a puzzling case among the extensively studied frustrated compounds of the rare-earth pyrochlore family. Indeed, while a first order transition towards a long-range antiferromagnetic state with the so-called Palmer-Chalker structure is theoretically predicted, it has not been observed yet, leaving the issue, as to whether it is a spin-liquid candidate, open. We report on neutron scattering and magnetization measurements which evidence a second order transition towards this Palmer-Chalker ordered state around 108 mK. Extreme care was taken to ensure a proper thermalization of the sample, which has proved to be crucial to successfully observe the magnetic Bragg peaks. At the transition, a gap opens in the excitations, superimposed on a strong quasielastic signal. The exchange parameters, refined from a spin wave analysis in applied magnetic field, confirm that \ersn\ is a realization of the dipolar XY pyrochlore antiferromagnet. The proximity of competing phases and the strong XY anisotropy of the \er\ magnetic moment might be at the origin of enhanced fluctuations, leading to the unexpected nature of the transition, the low ordering temperature, and the observed multi-scale dynamics.
\end{abstract}

\maketitle
Frustration in magnetism is usually characterized by the inability of a system to condense into an ordered state, even well below the temperature range of the magnetic interactions \cite{Lacroix}. This reflects the presence, at the classical level, of a large ground state degeneracy, which prevents the system from choosing a unique ground state. Nevertheless, the system may eventually order owing to the presence of additional terms in the Hamiltonian, like second neighbor or Dzyaloshinski-Moriya interactions, or owing to ``order by disorder" phenomena \cite{Villain80, Shender82}, which lift the degeneracy and stabilize a unique ordered state. Conversely, fluctuations, originating for example from the proximity of competing phases, can hinder magnetic ordering, resulting in an unconventional correlated state with exotic excitations.

\begin{figure}[!h]
\includegraphics[width=8cm]{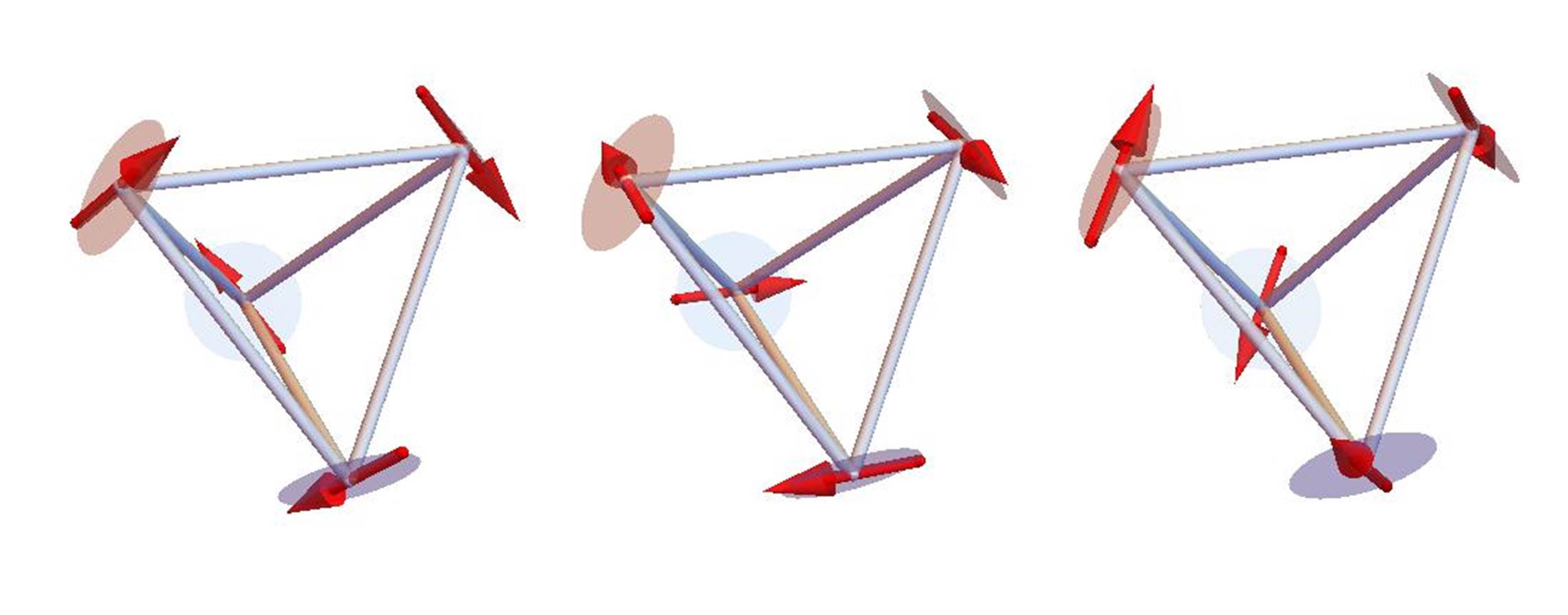}
\caption{(Color online) Sketch (on a single tetrahedron) of three Palmer-Chalker configurations forming distinct magnetic domains below $T_{\rm N}$ (three others are obtained by reversing all spins). 
The local XY planes perpendicular to the $\langle111\rangle$ axes are indicated by colored disks. 
Spins are pairwise anti-parallel, and collinear with an edge of the tetrahedron. These configurations can be described as chiral spin crosses. 
}
\label{fig_PC}
\end{figure}

The pyrochlore oxide \ersn\ appears to belong to this category. In this compound, the \er\ magnetic moments reside on the vertices of a lattice made of corner sharing tetrahedra, and are confined by a strong XY anisotropy within local planes, perpendicular to the $\langle111\rangle$ axes. The magnetic interactions are found to be governed by dipolar interactions, in addition to a quasi-isotropic antiferromagnetic exchange tensor \cite{Guitteny13}. Mean field calculations show that, in the $T=0$ phase diagram, this interaction tensor locates \ersn\ in an antiferromagnetic phase, called ``Palmer-Chalker" phase \cite{Palmer00} (see Figure~\ref{fig_PC}), close to the boundary with another ordered phase, the so-called $\psi_2$ phase, which is realized in the related compound \erti\ \cite{Poole07}. The predicted N\'eel temperature is about $T_{\rm N}^{\rm MF} \approx1.3$~K. Monte-Carlo simulations \cite{Yan13} have pointed out that fluctuations tend to lower the ordering temperature, with $T_{\rm N}^{\rm MC} \approx 200$~mK.

Experimentally, no phase transition has been detected  in \ersn\ down to 20~mK in muon spin relaxation measurements \cite{Lago05} and down to about 100 mK in magnetization \cite{Matsuhira02, Guitteny13} and neutron scattering experiments \cite{Sarte11, Bramwell04}. Yet Palmer-Chalker like short-range correlations have been reported below 5~K, taking the form of a broad diffuse quasielastic signal in neutron scattering measurements \cite{Sarte11, Guitteny13}. In this letter, using neutron diffraction and magnetization measurements, we show that \ersn\ does order in the Palmer-Chalker state at a N\'eel temperature $T_{\rm N} \approx 108$ mK. This long range ordering is characterized by magnetic Bragg peaks which develop on top of the broad diffuse scattering. The latter disappears progressively as the ordered magnetic moment increases, indicating a second order transition. Concomitantly, the slow dynamics previously observed in ac susceptibility above the magnetic transition persist at low temperature and coexist with the Palmer-Chalker ordering. In addition, on entering the ordered phase, inelastic neutron scattering (INS) experiments reveal new features in the spin excitation spectrum stemming from the magnetic Bragg peaks.
Using INS measurements performed in applied magnetic field, we determine the exchange parameters of a model Hamiltonian, confirming previous estimations.

Magnetization and ac susceptibility measurements were performed on a SQUID magnetometer equipped with a dilution refrigerator developed at the Institut N\'eel \cite{Paulsen01}. Neutron diffraction experiments were performed on the G4.1 diffractometer (LLB-Orph\'ee facility) with $\lambda=2.426$~\AA. INS measurements were carried out on the triple axis spectrometer 4F2 using a final wavevector $k_f=1.15$~\AA$^{-1}$. The energy resolution was about 70 $\mu$eV. The sample was a pure polycrystalline \ersn\ compound, synthesized by a solid state reaction from a stoichiometric mixture of Er$_2$O$_3$ (99.9\%) and SnO$_2$ (99.996\%). The powder was ground and heated for 6--8 h four times from 1400 to 1450$^{\circ}$C in air, cooled down to room temperature, and reground after each calcination. 

An important issue regarding the measurements at very low temperature concerns the thermalization of the powder sample. For magnetization measurements, a few milligrams of \ersn\ were mixed with apiezon N grease in a copper pouch, to improve the thermal contact and reduce the thermalization time. For neutron scattering experiments, a dedicated vanadium cell was used in the dilution fridge. The cell was filled with He gas up to 40 bars. During the experiments, it became obvious that a non negligible heating was induced by the sample's activation in the neutron beam. This effect was all the more important when the incident flux was large. Special care was then taken to minimize this effect for a better temperature control. For INS measurements, the neutron flux was reduced with a lead attenuator. For diffraction measurements, short counting times (of about 30 min) were programmed, alternating with deactivation (hence cooling) periods of 1 h. This thermalization issue probably explains why the transition had not been reported in previous neutron scattering measurements. 

\begin{figure}[!]
\includegraphics[width=8cm]{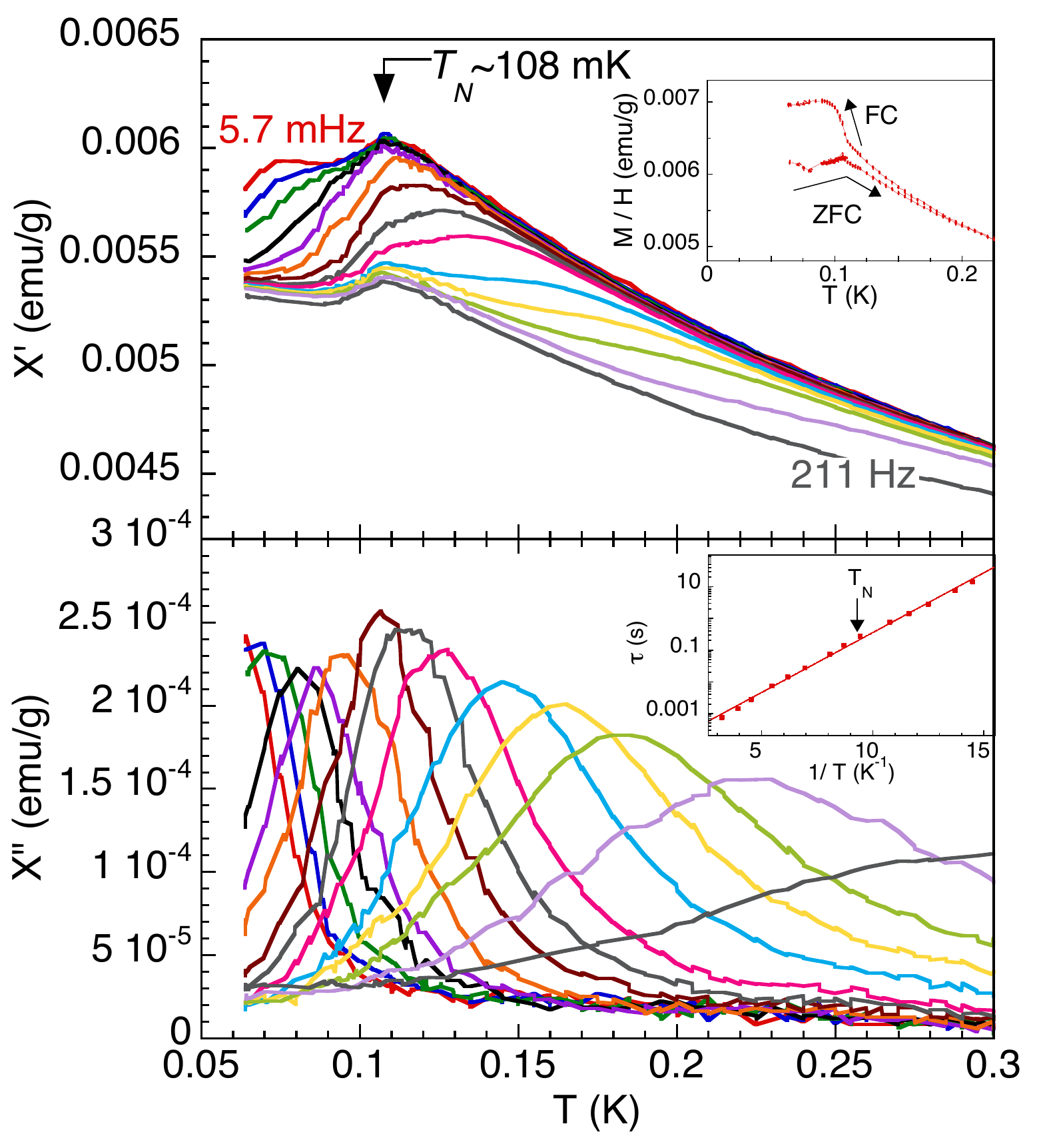}
\caption{(Color online) Ac susceptibility $\chi'$ (top) and $\chi''$ (bottom) vs temperature $T$ measured for several frequencies $f$ between 5.7 mHz and 211 Hz, with $\mu_0 H_{\rm ac}=0.1$~mT. Top inset: Magnetization $M/H$ vs $T$ measured in a ZFC-FC procedure with $\mu_0 H=1$ mT. Bottom inset: Relaxation time $\tau=1/2\pi f$ vs $1/T$ extracted from the maxima of $\chi"$. The line is a fit with the Arrhenius law: $\tau=\tau_0 \exp(E/T)$ with $\tau_0 \approx 6 \times 10^{-5}$ s and $E\approx 0.9$ K.
}
\label{fig_Xac-M}
\end{figure}

Zero field cooled - field cooled (ZFC-FC) magnetization measurements show an antiferromagnetic transition at $T_{\rm N}=108\pm5$ mK (see inset of Figure \ref{fig_Xac-M}). It manifests as a peak in the ZFC curve, while the FC magnetization sharply increases at the transition. This effect, although less pronounced, is similar to what is observed in \erti\ \cite{Petrenko11} and could be due to the polarization of uncompensated magnetic moments at domain boundaries. This magnetic transition is also evidenced by a peak at $T_{\rm N}$ in the real part of the ac susceptibility (see top of Figure \ref{fig_Xac-M}). 
A frequency dependent signal, a signature of slow dynamics, comes on top of this peak, as a bump which moves towards high temperature when the frequency increases. It is associated with a peak in the imaginary part of the ac susceptibility, which follows the same thermal activated law above and below the N\'eel temperature (see bottom inset of Figure~\ref{fig_Xac-M}), showing that the slow dynamics is not affected by the transition.

\begin{figure}[!]
\includegraphics[width=8cm]{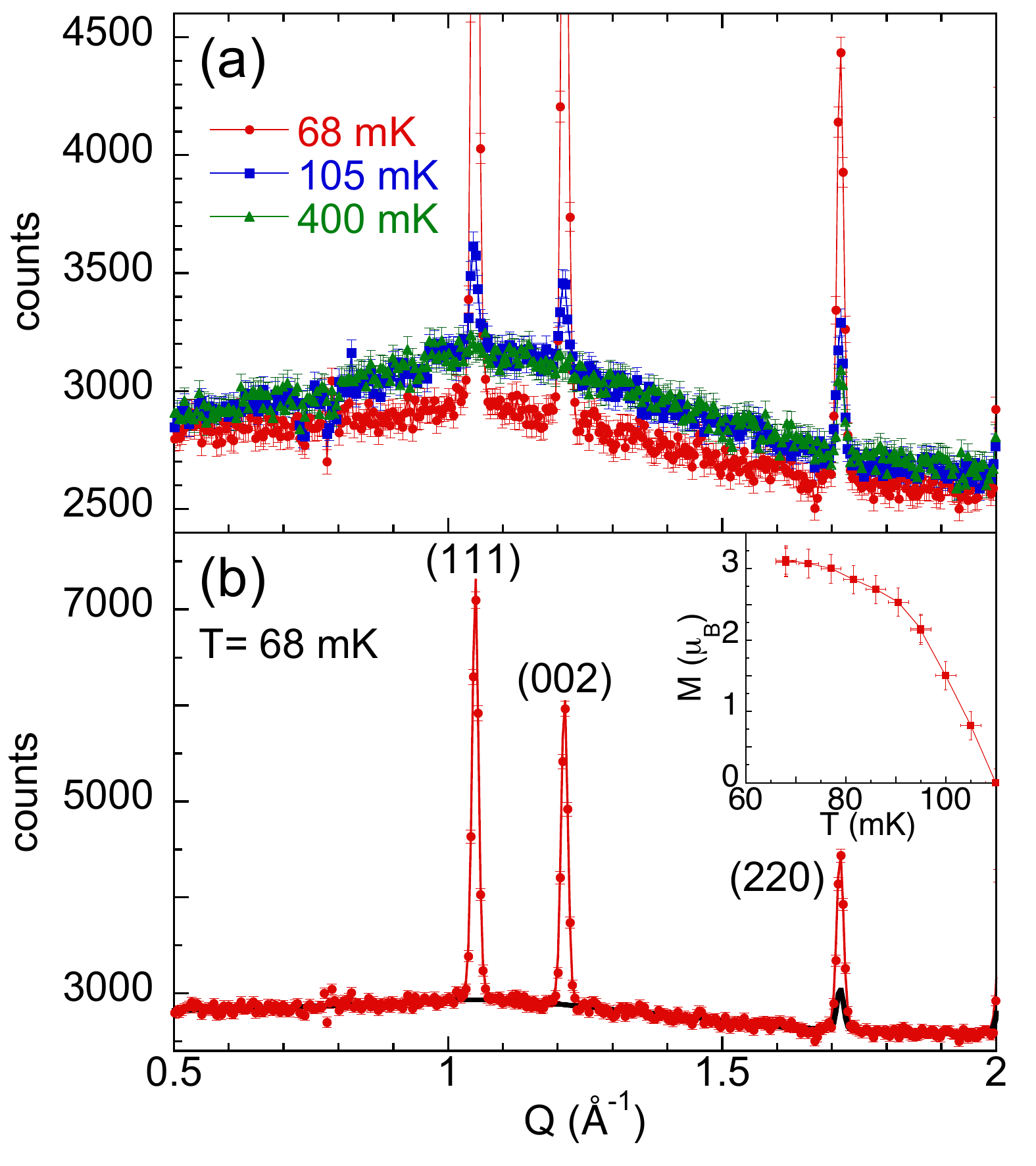}
\caption{(Color online) (a) Diffractograms at 68~mK (red), 105~mK (blue) and 400~mK (green). (b) Diffractogram at 68~mK. The black line is the fit corresponding to the nuclear contribution plus the diffuse Palmer-Chalker scattering. The red line shows the fit of the whole signal, including the magnetic long range Palmer-Chalker contribution. Inset: Temperature dependence of the ordered magnetic moment deduced from the refinements. 
}
\label{fig_diffraction}
\end{figure}

Neutron diffraction confirms the presence of antiferromagnetic ordering at very low temperature, as shown in Figure \ref{fig_diffraction}. Intensity increases on existing crystalline Bragg peak positions, indicating a ${\bf k}={\bf 0}$ propagation vector. The main magnetic peaks appear at $1.07$ and $1.22$~\AA$^{-1}$, corresponding to the $Q=(111)$ and $(002)$ wavevectors, and emerge from the diffuse scattering signal characteristic of short-range Palmer-Chalker correlations \cite{Sarte11, Guitteny13}. The latter has almost disappeared at the lowest temperature ($T=68$ mK), indicating that most of the magnetic moment is ordered (see Figure \ref{fig_diffraction}(a)). Rietveld refinements show that, among the possible irreducible representations authorized by the ${\bf k}={\bf 0}$ propagation vector \cite{Wills06, Poole07}, neutron intensities can only be properly modeled by the $\Gamma_7$-Palmer-Chalker  representation \cite{supmat}. At the lowest temperature, the ordered moment is $3.1~\mu_{\rm B}$, i.e. about 80 \% of the magnetic moment of the \er\ ground state doublet, estimated at $3.8~\mu_{\rm B}$ \cite{Guitteny13} (see Figure \ref{fig_diffraction}(b)). The remaining \er\ moments are embedded in short-range only Palmer-Chalker correlations, as reflected by the persistence of a weak diffuse signal. This ordered moment value is consistent with what is expected in a conventional second order phase transition at the corresponding $T/T_{\rm N}$ ratio. The second order nature is further confirmed by the gradual increase of the ordered moment below $T_{\rm N}$ (see inset of Figure \ref{fig_diffraction}(b)). 

\begin{figure*}[!]
\includegraphics[width=\textwidth]{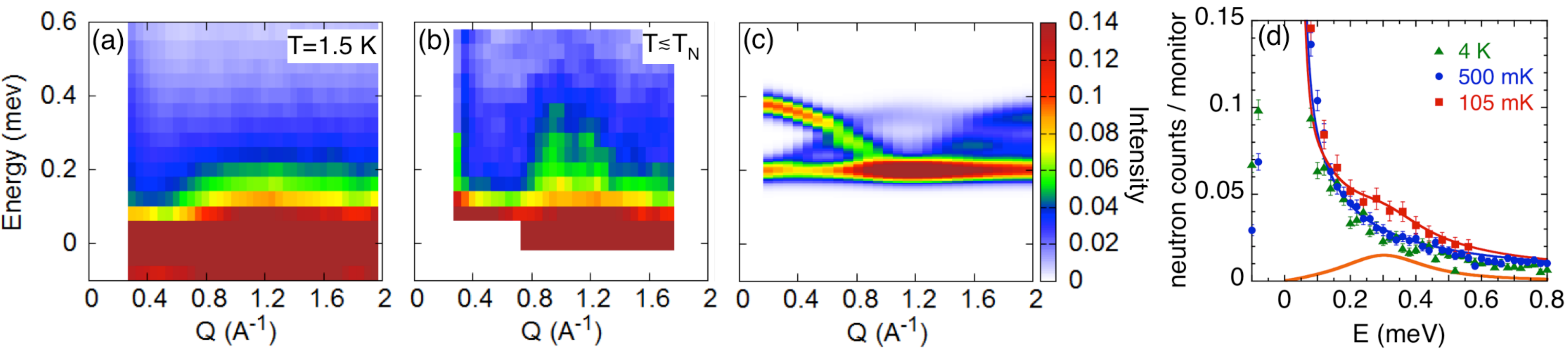}
\caption{(Color online) Inelastic neutron scattering in zero field: powder averaged spectrum $S(Q,\omega)$ measured (a) at 1.5~K, (b) below $T_{\rm N}$, at about 105 mK. The neutron intensity is normalized to the monitor. Constant energy-scans were carried out to minimize self-heating effects. (c) Excitation spectrum in the Palmer-Chalker phase obtained from RPA calculations with the exchange parameters ${\cal J}_a={\cal J}_b=0.03$~K, ${\cal J}_c=0.05$ K and ${\cal J}_4=0$ at $T=0$. (d) Energy cuts averaged for $1 \leq Q \leq 1.2$ \AA$^{-1}$ at several temperatures: 4 K (green triangles), 500 mK (blue dots), 105 mK (red squares). The blue line is a fit involving a Gaussian profile, centered at zero energy to model the elastic response, and a quasielastic contribution $(1+n(\omega)) \times A \omega \Gamma /(
\omega^2+\Gamma^2)$. $1+n(\omega)$ is the detailed balance factor, $A=0.2$ and $\Gamma=$ 0.046~meV. The red line contains an additional inelastic contribution, shown separately by the orange line, and described by a Lorentzian profile $(1+n(\omega)) \times B \gamma(1/[(\omega-\omega_o)^2+\gamma^2]-1/[(\omega+\omega_o)^2+\gamma^2])$ with $B=0.016$, $\omega_o=0.3$~meV and $\gamma=0.016$ meV.}
\label{fig_INS2}
\end{figure*}

We now focus on the excitation spectrum associated with this magnetic ordering. As previously reported, at temperatures as high as 10~K, the INS spectrum includes a strong quasielastic signal \cite{Sarte11}. Its intensity is stronger around $Q_o=1.1$ \AA$^{-1}$, which corresponds to the position of the maximum intensity of the Palmer-Chalker diffuse scattering (see Figure \ref{fig_INS2}(a)). On cooling, the intensity increases, while the width decreases, corresponding to a slowing down of the fluctuations, reaching a characteristic time of about $10^{-11}$~s close to $T_{\rm N}$. This quasielastic signal persists down to temperatures below $T_{\rm N}$ (see Figure \ref{fig_INS2}(b)). Nevertheless, because of self-heating in the neutron beam, the effective temperature actually reached by the sample at the lowest temperature of the dilution fridge ($T_{\rm min}=70$ mK), is estimated to be 105 mK from the intensity of the magnetic Bragg peaks, thus just below $T_{\rm N}$ (corresponding to an ordered moment of $0.8~\mu_{\rm B}$).  Additional features, stemming from the Bragg peaks at $Q\approx1$ and $1.2$~\AA$^{-1}$, also arise in the ordered regime (see Figure \ref{fig_INS2}(b)). In the cut shown in Figure \ref{fig_INS2}(d), this additional signal manifests as a broad band in the $0.2 \leq \omega \leq 0.5$ meV range, superimposed on the quasielastic response. 

\begin{figure}[!t]
\includegraphics[width=8.5cm]{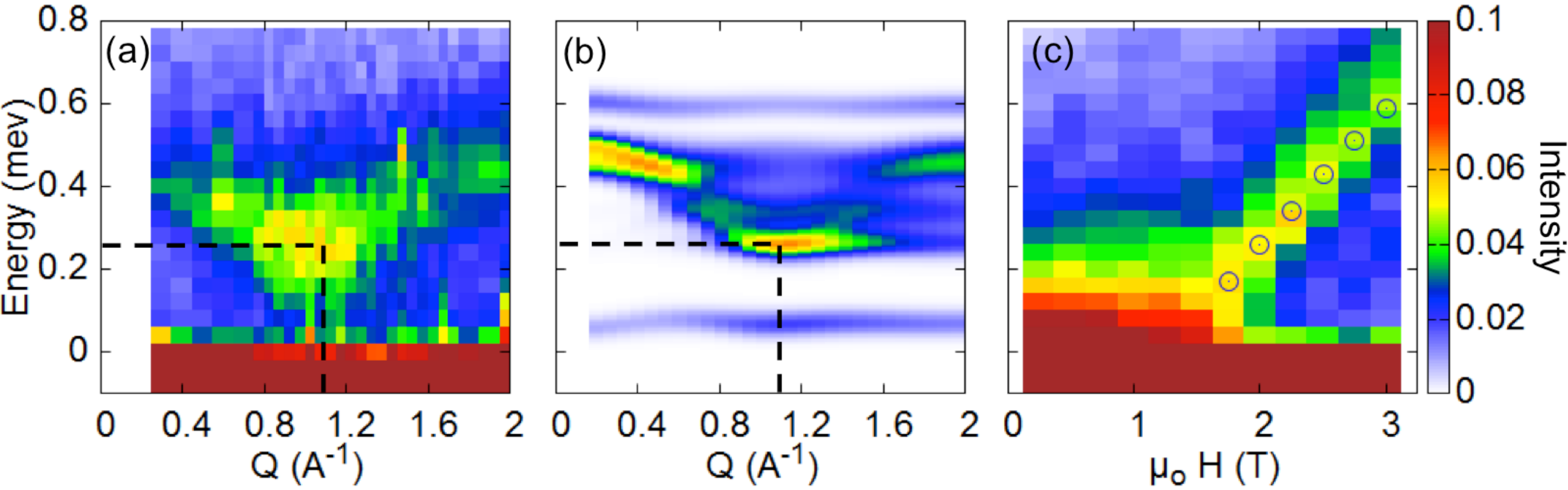}
\caption{(Color online)
Inelastic neutron scattering in the field polarized phase. (a) Powder average spectrum $S(Q,\omega)$ measured at $T=1.5$ K with $\mu_0 H=2$ T. (b) Calculated powder average spectrum $S(Q,\omega)$ at 2 T and $T=0$ with ${\cal J}_a={\cal J}_b=0.03$ K, ${\cal J}_c=0.05$ K and ${\cal J}_4=0$. (c) Powder average spectrum at $Q_o=1.1$ \AA$^{-1}$ and at $T=1.5$ K as a function of field $H$. The empty blue circles are the values of the gap obtained from the calculations with the above parameters. The neutron intensity is normalized to the monitor.}
\label{fig_INS_H}
\end{figure}

To further analyze these results, we have refined the exchange parameters previously estimated from magnetization measurements \cite{Guitteny13}. 
We have used the well-documented procedure \cite{Savary12, Ross11}, which consists in applying a magnetic field to drive the ground state towards a field polarized state and analyzing the spin excitations in terms of conventional spin waves. Owing to the polycrystalline nature of the \ersn\ sample, the magnetic field is applied simultaneously in all crystallographic directions, leading to an average excitation spectrum. INS measurements were performed as a function of magnetic field at 1.5 K. Above about $\mu_o H = $ 1.5 T, the response is no longer quasielastic-like, and a dispersive spectrum is observed, with the opening of a gap (see Figure \ref{fig_INS_H}(a)). Its value is $\Delta = 0.26 \pm 0.04$~meV at $Q_o =1.1$~\AA$^{-1}$ for a field of $\mu_o H = 2$ T. $\Delta$ increases roughly linearly further increasing the field, as illustrated in Figure \ref{fig_INS_H}(c).

To determine the exchange parameters, we consider the following Hamiltonian: 
$
{\cal H} = {\cal H}_{\rm CEF} + 
\frac{1}{2}\sum_{<i,j>} {\bf J}_i {\cal \tilde J}{\bf J}_j + g_{\rm J} \mu_{\rm B} { \bf H}. {\bf J}_i
$, where
$\bf{J}$ is the \er\, magnetic moment, ${\cal H}_{\rm CEF}$ is the crystal electric field (CEF) Hamiltonian, $\bf{H}$ denotes the magnetic field, $g_{\rm J}$ is the effective $g$ factor, and ${\cal \tilde J}$ is the anisotropic exchange tensor, which incorporates the dipolar interaction truncated to its nearest neighbors. It is written in the $(\bf{a}_{ij},\bf{b}_{ij},\bf{c}_{ij})$ frame linked with \er-\er $\langle ij \rangle$ nearest neighbors bonds \cite{Malkin10}:
\begin{align*}
{\bf J}_i {\cal \tilde J} {\bf J}_j = &\sum_{\mu,\nu=x,y,z} J_i^{\mu} D_{\rm nn} \left(a_{ij}^{\mu} a_{ij}^{\nu} + b_{ij}^{\mu} b_{ij}^{\nu} -2 c_{ij}^{\mu} c_{ij}^{\nu} \right) J_j^{\nu} \\
+&\sum_{\mu,\nu=x,y,z}
 J_i^{\mu} \left({\cal J}_a a_{ij}^{\mu} a_{ij}^{\nu} + {\cal J}_b b_{ij}^{\mu} b_{ij}^{\nu} + {\cal J}_c c_{ij}^{\mu} c_{ij}^{\nu} 
\right) J_j^{\nu} \\
 +& {\cal J}_4 \sqrt{2}~{\bf b}_{ij}.({\bf J}_i \times {\bf J}_j)
\end{align*}
where $D_{\rm nn} = \dfrac{\mu_o}{4\pi}\dfrac{(g_{\rm J} \mu_{\rm B})^2}{r^3_{\rm nn}}=0.022$ K is the pseudo-dipolar contribution (with $r_{\rm nn}$ the nearest neighbor distance in the pyrochlore lattice), ${\cal J}_a$, ${\cal J}_b$, ${\cal J}_c$ are effective exchange parameters, and ${\cal J}_4$ corresponds to the antisymmetric Dzyaloshinski-Moriya interaction. This model takes into account the specific CEF scheme which strongly confines the spins within the local XY planes \cite{Guitteny13}. The powder average of the spin excitation spectrum $S(Q,\omega)_H$ is then calculated for a given field ${\bf H}$ using the random phase approximation (RPA).

To compare with the polycrystalline experimental data shown in Figure \ref{fig_INS_H}(a), the average over all field directions has to be performed. As a first approximation, we consider that the powder spectrum can be described by averaging over the high symmetry directions of the system, $\langle001\rangle$, $\langle110\rangle$ and $\langle111\rangle$, taking into account their multiplicity: 
 $\overline{S(Q,\omega)} =\frac{1}{13}( 3 S(Q,\omega)_{{\bf H} \parallel [001]}+6 S(Q,\omega)_{{\bf H} \parallel [110]}+4 S(Q,\omega)_{{\bf H} \parallel [111]})$. Assuming $J_4=0$, since the Dzyaloshinski-Moriya is expected to be small compared to symmetric exchange, the best agreement is obtained with ${\cal J}_a={\cal J}_b=0.03 \pm 0.003$~K, ${\cal J}_c=0.05 \pm 0.01~$K, thus consistent with the uncertainty range given in Ref. \onlinecite{Guitteny13}. Calculations reproduce the strong intensity close to $Q_o$ and the presence of a gap of about 0.25~meV (see Figure \ref{fig_INS_H}(b)). They also account for the field dependence of the gap at $Q_o$ above 1.5 T, as shown in Figure \ref{fig_INS_H}(c) by the blue open circles. 
RPA calculations performed with these exchange parameters at $T=0$ in zero field, predict a gapped flat mode at 0.2 meV, together with excitations dispersing up to 0.4 meV (see Figure \ref{fig_INS2}(c)). While the value of the gap is consistent with the measured inelastic contribution, the predicted dispersion is hardly distinguishable in the data of Figure~\ref{fig_INS2}(b). 

It is instructive to compare the \ersn\ behavior with results obtained on \gdsn. The latter, in which the \gd\ magnetic moment is almost isotropic, is known as the archetype of the dipolar {\it Heisenberg} pyrochlore antiferromagnet. In spite of very different anisotropies in both compounds, calculations predict identical behaviors, for thermodynamic and dynamical properties: the transition towards the Palmer-Chalker phase is expected to be first order \cite{Cepas05, Yan13}, while a spin wave spectrum, similar to Figure \ref{fig_INS2}(c), should develop at low temperature, consisting in dispersive branches on top of a gapped flat mode \cite{DelMaestro04, DelMaestro07, Sosin09}. \gdsn\ indeed follows these predictions \cite{Bonville03, Stewart08}, and the opening of the gap occurs well above $T_{\rm N}$ \cite{Sosin08}. In \ersn, the scenario appears more complex: the temperature dependence of the ordered magnetic moment points out a second order transition. 
The lack of clear dispersion in the experimental INS spectrum can be attributed to the proximity of $T_{\rm N}$, so that conventional spin waves are expected to develop at lower temperature. However, the strong XY character of the system, which can lead to unconventional excitations as lines of vortices \cite{Savit1978}, insufficiently captured by the RPA, might induce a peculiar $Q$ dependence of the spectrum.

The strong ratio between dipolar and exchange interactions in \ersn, which is about $0.5 - 0.7$ (against 0.15 in \gdsn\ \cite{DelMaestro07}) also appears crucial. Preliminary calculations with \ersn\ parameters at $T=0$ and ${\bf Q}=(1,1,1)$ show that accounting for the long range part of the dipolar interaction beyond the truncation considered above and in Ref.~\onlinecite{Guitteny13, Yan13}, slightly reduces the gap value, but does not affect the ground state. It was however established in Heisenberg systems that, at finite temperature, long range dipolar interactions tend to weaken the Palmer-Chalker state stability, due to the proximity of an unconventional state with a ${\bf k}=(1/2, 1/2, 1/2)$ propagation vector \cite{Enjalran03, Cepas04} (proposed to be the magnetic state stabilized in \gdti\ \cite{Champion01}). The proximity of \ersn\ with such a state, in addition to the proximity with the $\psi_2$ state evoked above, might reinforce the fluctuations in the Palmer-Chalker state at finite temperature and explain the low ordering temperature, the unexpected second order of the transition as well as the structure of the spectrum close to the transition. Note that the very low temperature considered here could suggest a role of quantum fluctuations, but they have been shown not to destabilize the Palmer-Chalker state much \cite{DelMaestro04, Yan13}.
 
Finally, as reported in numerous frustrated magnets \cite{Gardner10}, \ersn\ hosts a complex spin dynamics where   several time scales coexist, even in the long-range order regime. In addition to the fast fluctuations observed by inelastic neutron scattering, slow dynamics are observed in ac susceptibility above $T_{\rm N}$. They persist below $T_{\rm N}$, coexisting both with the magnetic ordering along with magnetic moments fluctuating at the muon time scale \cite{Lago05}. The characteristic energy barrier of this slow dynamics, $E=0.9$~K ($=0.078$ meV), is smaller than the gap of the Palmer-Chalker state. It might be associated with the spins at the boundary between the six existing domains, as previously reported in kagome and pyrochlore systems \cite{Lhotel11, Tardif15}.

We have shown that the pyrochlore \ersn\ orders in the Palmer-Chalker state at about 100 mK. As confirmed by the analysis of the exchange couplings, it is the realization of the dipolar XY pyrochlore antiferromagnet. The absence of a first order transition along with multi-scale dynamics are signatures of an unconventional magnetic state. Further calculations, accounting for the XY character and for the long range dipolar interactions at finite temperature are needed to understand the role of fluctuations and may enlighten the existence of exotic excitations. The proximity of \ersn\ with competing phases might also enhance fluctuations, as recently proposed in the context of another pyrochlore compound \ybti\ \cite{Robert15, Jaubert15}. In that perspective, our study points out the importance of multiphase competitions in the novel magnetic states emerging in frustrated systems.

\acknowledgements
{\it Acknowledgements:} We thank C. Paulsen for allowing us to use his SQUID dilution magnetometers. This work was partly funded by the ANR {\it Dymage} 13-BS04-0013 and the {\it 1Dmag} PALM project. 


\vspace{2 cm}

\renewcommand{\figurename}{FIG. S}
\renewcommand{\tablename}{TABLE S}
\renewcommand{\theequation}{\alph{equation}}


 \setcounter{figure}{0} 

\begin{center}
{\bf {\large Supplemental Material}}
\end{center}

\bigskip


This document describes additional information concerning the refinement of the ${\bf k}=0$ magnetic structure in the \rt\ pyrochlore magnets. \\

As a preamble, let us recall that in the pyrochlore structure, the relevant rare earth magnetic ions are located on the vertices of corner sharing tetradehra. It is then convenient to introduce site dependent $({\boldsymbol a}_i,{\boldsymbol b}_i,{\boldsymbol z}_i)$ local frames associated with the four sites of a tetrahedron (see Table S\ref{table1}). The ${\boldsymbol z}_i$ denote the local crystal field (CEF) axis.\\

\begin{table}[h]
\setlength{\extrarowheight}{1pt}
\begin{tabularx}{8.5cm}{M{1.75cm}M{1.75cm}YM{1.5cm}Y} 
\hline
\hline
Site & 1 & 2 & 3 & 4 \\
CEF axis ${\boldsymbol z_i}$ & $(1,1,\bar{1})$ & $(\bar{1},\bar{1},\bar{1})$ & $(\bar{1},1,1)$ & $(1,\bar{1},1)$ \\
Coordinates & {\text{\footnotesize$(1/4,1/4,1/2)$}} & {\text{\footnotesize$(0,0,1/2)$}} & {\text{\footnotesize$(0,1/4,3/4)$}} & {\text{\footnotesize$(1/4,0,3/4)$}}\\
\hline 
${\boldsymbol a_i}$ & $(\bar{1},\bar{1},\bar{2})$ & $(1, 1,\bar{2})$ & $(1,\bar{1},2)$ & $(\bar{1},1,2)$ \\
${\boldsymbol b_i}$ & $(\bar{1},1,0)$ & $(1, \bar{1},0)$ & $(1,1,0)$ & $(\bar{1},\bar{1},0)$ \\
\hline
\hline
\end{tabularx}\caption{Coordinates, written in the cubic $Fd\bar{3}m$ structure of the pyrochlore lattice, of the site dependent local ${\boldsymbol a_i}$ and ${\boldsymbol b_i}$ vectors spanning the local XY anisotropy planes. The CEF axes ${\boldsymbol z_i}$ of the rare earth is perpendicular to those planes. }
\label{table1}
\end{table}

Different ${\bf k}=0$ magnetic structures have been reported in literature. They are stabilized in various pyrochlore magnets. Note that, importantly, the 4 tetrahedra that built the unit cell are always identical. Owing to the $Fd\bar{3}m$ crystallographic space group, and according to Reference \onlinecite{Wills06} for example, the different magnetic configurations can be classified in terms of the irreducible co-representations listed below:\\

\begin{enumerate} 
\item $\Gamma_3$ corresponds to the ``all in -- all out'' structure, as in FeF$_3$ \cite{Ferey86} or in \ndzr\  \cite{Lhotel15} for instance. Here, all spins point out or into a given tetrahedron. The spins $\left\{{\bf S}_{1,2,3,4}\right\}$ are then given by 
$${\bf S}_i= {\boldsymbol z}_i$$

\item $\Gamma_5$ describes a structure where the spins show the same $\phi$ angle within the local XY planes: $${\bf S}_i= \cos\phi ~{\boldsymbol a}_i+\sin\phi ~{\boldsymbol b}_i.$$ This representation is spanned by two basis vectors labeled $\psi_2$ and $\psi_3$ which correspond respectively to $\phi=0$ and $\phi=\pi/2$. The ground state of \erti\ is, for instance, described by the $\psi_2$ configuration \cite{Poole07}.

\item $\Gamma_7$ is the so-called Palmer-Chalker structure (after the name of the two physicists) where the spins are given by : 
\begin{eqnarray*}
{\bf S}_{1,4} &=& {\boldsymbol b}_{1,4}\\
{\bf S}_{2,3} &=& -{\boldsymbol b}_{2,3}
\end{eqnarray*}
This situation describes the ground state in \gdsn\ \cite{Wills06} and, as we show in this work, in \ersn. Note that 3 domains can be defined.

\item $\Gamma_9$ is a splayed ferromagnetic structure, akin to a ``two in -- two out'' structure (where two spins point into and 2 out of a given tetrahedron) but with an additional component along one of the cubic axis. This is the case for instance in \ybsn\ \cite{Yaouanc13}. In this case, the structure is defined as: 
\begin{eqnarray*}
{\bf S}_{1,2} &=& \lambda {\boldsymbol z}_{1,2} + \mu {\boldsymbol c}  \\
{\bf S}_{3,4} &=& - \lambda {\boldsymbol z}_{3,4} + \mu {\boldsymbol c}
\end{eqnarray*}
where $\lambda$ and $\mu$ determine the tilt angle out of the CEF axis.
\end{enumerate} 

\begin{figure}[t]
\includegraphics[width=8.5cm]{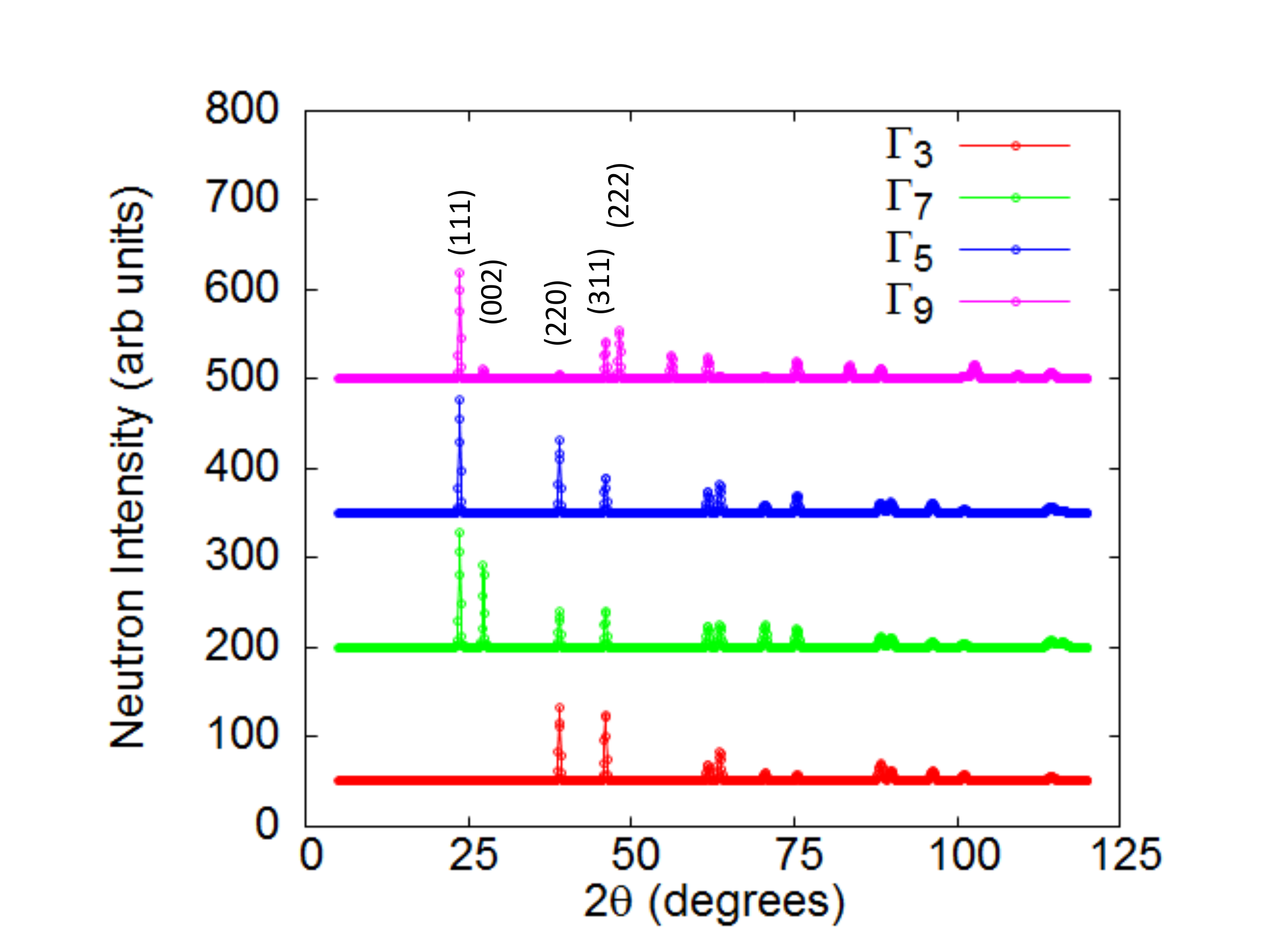}
\caption{Magnetic diffraction pattern for the different $\Gamma_i$ configurations listed in the text. The calculations are performed for a magnetic moment of 1 $\mu_{\rm B}$ and a wavelength of 2.4 \AA. The results shown for $\Gamma_9$ correspond to the specific case of the \ybsn\ magnetic structure.  Note that the nuclear contribution is not shown. The patterns are shifted artificially for the sake of clarity. }
\label{fig1_supmat}
\end{figure}
Since these structures are ${\bf k}=0$ ones, the magnetic intensity appears on Q-positions where a nuclear scattering is also present. This makes the refinement {\it a priori} difficult, leaving only a few reflections, especially at small Q wavevectors, to disentangle the magnetic and nuclear contributions, and finally to refine the magnetic structure. Indeed, the nuclear signal is usually strong at large Q wavevectors, while at the same time, the magnetic scattering decreases because of the magnetic form factor. The refinement thus relies solely on these small Q wavevectors. In some cases, taking the subtraction of the high and low temperature data is the only way to isolate the magnetic contribution. Fortunately, in the present case, these different structures listed above give rise to quite different {\it magnetic} contributions, as shown in Figure S\ref{fig1_supmat}. This allows to identify the structure quite safely, even with only a few Bragg peak intensities. 
\newpage

Figure S\ref{fig1_supmat} shows the calculated magnetic patterns for those different cases. Clearly, they differ markedly from one to the other, leaving no room for doubt:  

- $\Gamma_3$ is easily identified since the intensities at ${\bf Q}=(111)$ and ${\bf Q}=(002)$ are zero. Furthermore the intensities of ${\bf Q}=(220)$ and ${\bf Q}=(311)$ are almost identical.

- $\Gamma_5$ and $\Gamma_7$ differ from the ${\bf Q}=(002)$ intensity (equals zero for $\Gamma_5$) and from the relative intensities of the ${\bf Q}=(111)$ and ${\bf Q}=(220)$ reflections.

- Finally $\Gamma_9$ in \ybsn\ is characterized by a very strong ${\bf Q}= (111)$ reflection and very small ${\bf Q}=(002)$ and  ${\bf Q}=(220)$ ones.

%
%
%
%
%
%

\end{document}